\documentclass[a4paper]{jpconf}
\usepackage{graphicx}
\usepackage{epsfig}
\usepackage{latexsym}
\usepackage{amssymb}
\usepackage{amsbsy}
\usepackage{amsmath}
\usepackage{cite}
\usepackage{epsfig}
\usepackage{graphics,color}
\usepackage{a4}
\newcommand{\as}{\alpha_s}

\newcommand{\be}{\begin{equation}}
\newcommand{\ee}{\end{equation}}

\newcommand{\bea}{\begin{eqnarray}}
\newcommand{\eea}{\end{eqnarray}}
\newcommand{\bean}{\begin{eqnarray*}}
\newcommand{\eean}{\end{eqnarray*}}

\newcommand{\half} {\frac{1}{2}}
\newcommand{\om} {\omega}

\newcommand{\pv}{{\mathbf p}}
\newcommand{\nv}{{\mathbf n}}

\begin{document}
\title{Bottomonium from lattice QCD as a probe of the Quark-Gluon Plasma}

\author{G.~Aarts, C.~Allton} 

\address{ Department of Physics, Swansea University, Swansea, United Kingdom} 

\author{A.~Kelly, J.-I.~Skullerud}

\address{Department of Mathematical Physics, National University of Ireland Maynooth,
Maynooth, County Kildare, Ireland}

\author{S.~Kim}
\address{Department of Physics, Sejong University, Seoul 143-747, Korea} 

\author{T.~Harris, S.~M.~Ryan} 
\address{School of Mathematics, Trinity College, Dublin 2, Ireland} 

\author{M.~P.~Lombardo}
\address{INFN-Laboratori Nazionali di Frascati, I-00044, Frascati (RM) Italy}

\author{M.~B.~Oktay}
\address{Physics Department, University of Utah, Salt Lake City,  Utah, USA} 

\author{D.~K.~Sinclair}
\address{ HEP Division, Argonne National Laboratory, 9700 South Cass Avenue,Argonne, Illinois 60439, USA}

\begin{abstract}
We study the temperature dependence of bottomonium for temperatures in the 
range $0.4 T_c < T < 2.1 T_c$,  using non-relativistic dynamics for the bottom 
quark and full relativistic lattice QCD simulations for $N_f=2$ light 
flavors.  We consider the behaviour of the
correlators in Euclidean space, we analyze the associated spectral functions
and we study the dependence on the momentum. 
Our results are amenable to a 
successful comparison with effective field
theories. They help build a coherent picture of the behaviour
of bottomonium in the plasma, consistent which the  current
LHC results. 
\end{abstract}

\section{Introduction} 
In this note
we review the results on the spectrum of bottomonium in the quark-gluon plasma 
obtained by our FASTSUM collaboration
\cite{Aarts:2010ek,Aarts:2011sm,Aarts:2012ka} 
using a hybrid approach which combines exact relativistic dynamics for the
light quarks with a non-relativistic approach for the bottom quark.

Recently, the first
 detailed results on the sequential Upsilon suppression
in PbPb collision at LHC energies have appeared \cite{:2012fr}
(see also several talks at this meeting \cite{BR}). 
These results call for a  quantitative understanding of the behaviour
of bottomonium at high temperature. 
Due to their considerable experimental and theoretical interest
these studies  have been carried out in the past 
with a variety of techniques. We refer to  recent excellent reviews
\cite{Hatsuda_QM,Kaczmarek:2012ne}, 
as well as to our published papers, for background material
and a complete set of references to early studies; here we limit ourselves to an exposition of our approach and 
results. 

In the next Section we  outline our approach to the
problem. Next,  we present the information we have 
 obtained from the correlators alone. We continue by showing the results
on the spectral functions. We finally discuss momentum  dependence
and  we close with a short summary.

\section{Five steps to Bottomonium}

Our strategy is  to treat as accurately as possible
the light quarks, and use non-relativistic QCD (NRQCD) for the $b$ quarks. 
In this way we hope to keep  the systematic effects under control.
Moreover, the accuracy of the results
can be improved following the standard lattice techniques. 
In short outline, 
this is our `five steps approach to bottomonium' strategy:

\noindent
{\bf i)}  Use full relativistic dynamics for up and down quarks 
(the inclusion of the 
strange quark is in progress).  Gauge configurations 
with two degenerate dynamical light Wilson-type quark flavors are produced on 
highly anisotropic lattices, 
in a range of temperatures comprised between $0.4T_c$
and $2.1T_c$. 
Details of the lattice action and parameters can be found in 
Refs.~\cite{Aarts:2007pk,Morrin:2006tf}. 

\noindent 
{\bf ii)} We use  NRQCD for the bottom quarks. 
Since NRQCD relies on the scale separation 
$M\gg T$ and we study temperatures up to $2 T_c \simeq 400$ MeV, its 
application in our temperature range 
is fully justified. The light quarks, instead, feel the full
relativistic dynamics. 
We computed NRQCD propagators on our 
configurations using a mean-field improved action with tree-level 
coefficients, which includes
terms up to and including ${\cal O}(v^4)$,
where $v$ is the typical velocity of a bottom quark in bottomonium.

\noindent
{\bf iii)} Analyze bottomonium correlators in Euclidean space. In standard
conditions we expect an exponential decay. An analytic model for the Euclidean
thermal behaviour is not known --- we contented ourselves to study 
what to expect when quarks are no longer bound. Consider 
free quarks in continuum NRQCD with energy $E_\pv=\pv^2/2M$. The 
correlators for the $S$ and $P$ waves are then of the form 
\cite{Burnier:2007qm}
 \begin{align}
\label{eq:GS}
 G_{S}(\tau)  \sim& \int \frac{d^3p}{(2\pi)^3}\, \exp(-2E_\pv\tau) 
  \sim \tau^{-3/2}, \\ 
\label{eq:GP}
 G_{P}(\tau)  \sim & \int \frac{d^3p}{(2\pi)^3}\, \pv^2 \exp(-2E_\pv\tau)  
  \sim \tau^{-5/2},
\end{align}
i.e., they decay as a power for large euclidean time. Of course,
interactions and finite lattice spacing and volume effects are expected
to modify this in the realistic case. Moreover, again, this is a limiting
high-temperature behaviour which should not be expected to hold even 
if the quarks unbind in the plasma. 

\noindent
{\bf iv)} Obtain spectral functions. In NRQCD, 
 the 
spectral relation  reduces to
 \be
 \label{eq:Gnr}
 G(\tau) =
 \int_{-2M}^\infty\frac{d\omega'}{\pi}\, \exp(-\om'\tau) \rho(\omega')
 \;\;\;\;\;\;(\mbox{NRQCD}), 
 \ee
 even at nonzero temperature
(note $\om=2M+\om'$). 
 As a result, all problems associated with
thermal boundary conditions are absent.
We will extract NRQCD spectral functions from the euclidean correlators 
using the Maximum Entropy Method (MEM) \cite{Asakawa:2000tr}.  

\noindent
{\bf v)} Extract physics from the results, discuss phenomenology,
validate effective models. We underscore that any meaningful 
comparison with experimental results and continuum physics requires
control over systematic and  lattice artifacts.

\begin{figure}[t]
 {\includegraphics*[width=6.5cm]{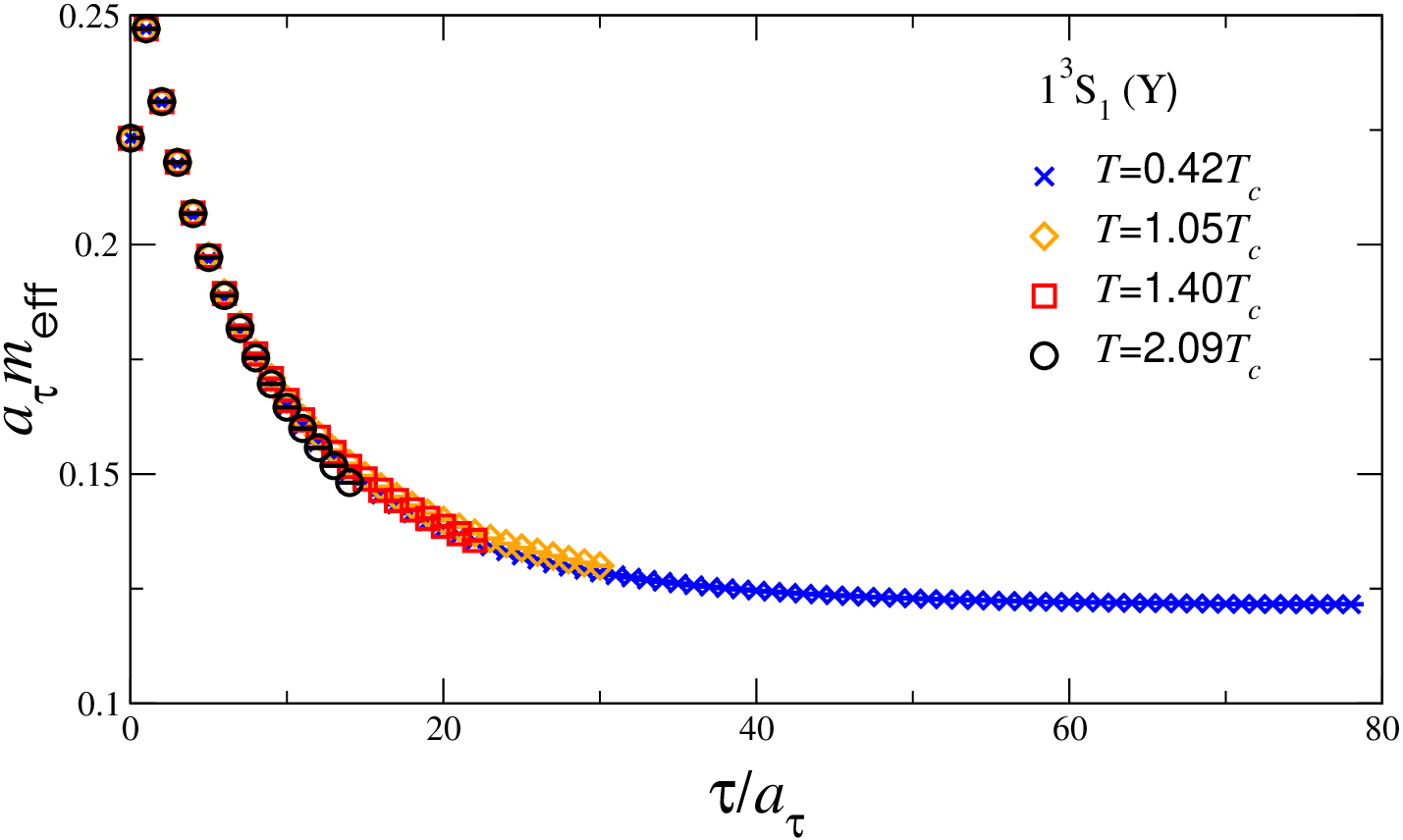}}
{\includegraphics*[width=6.5cm]{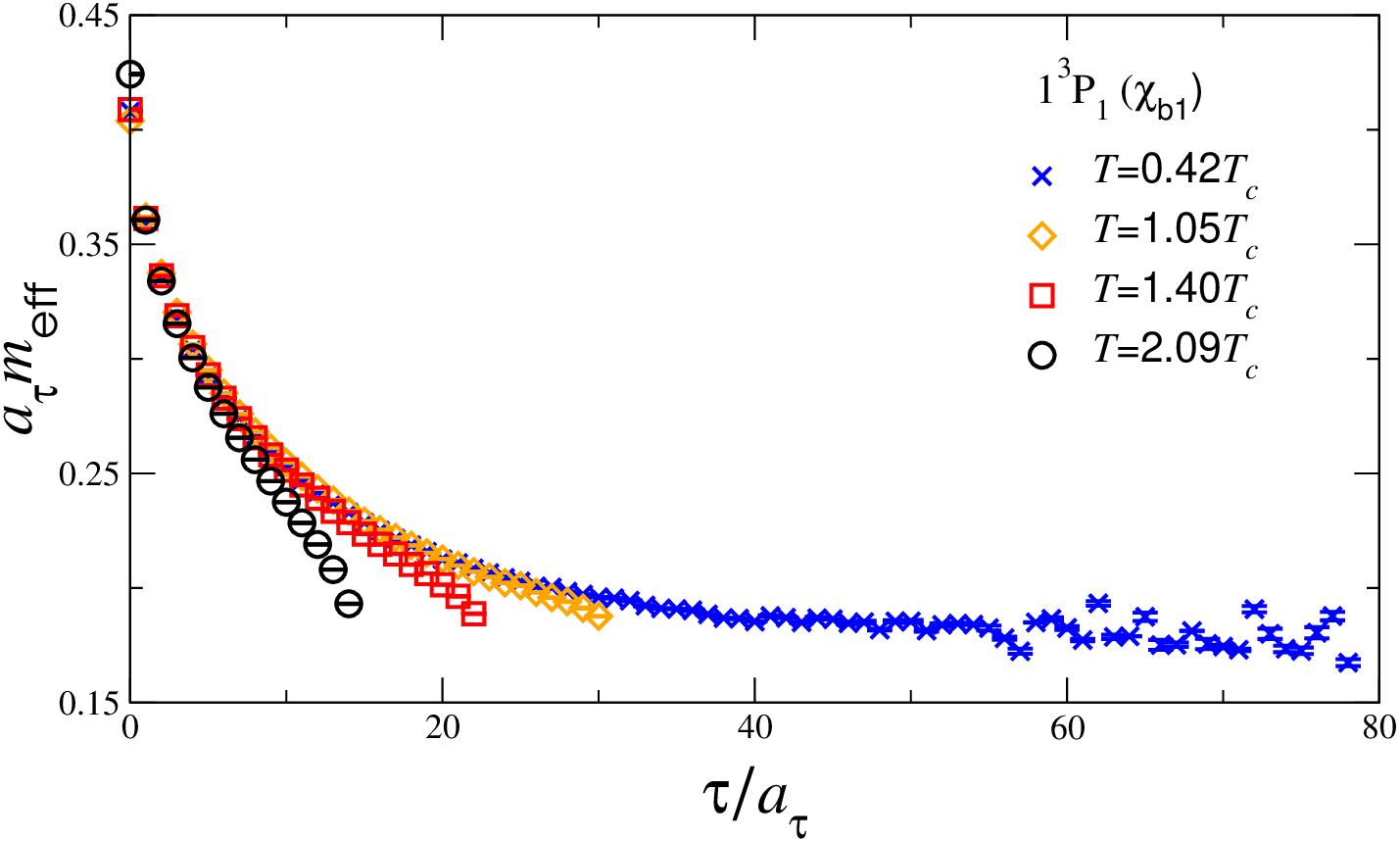}}
 \caption{
 Effective mass plots for the $\Upsilon$ (left) and  $\chi_{b1}$ 
(right) using point sources for various temperatures: note the different 
temperature dependence (Ref.\ \cite{Aarts:2010ek}).} 
 \label{fig:effmass}
 \end{figure}

 \begin{figure}[t]
 {\includegraphics[width=6.5cm]{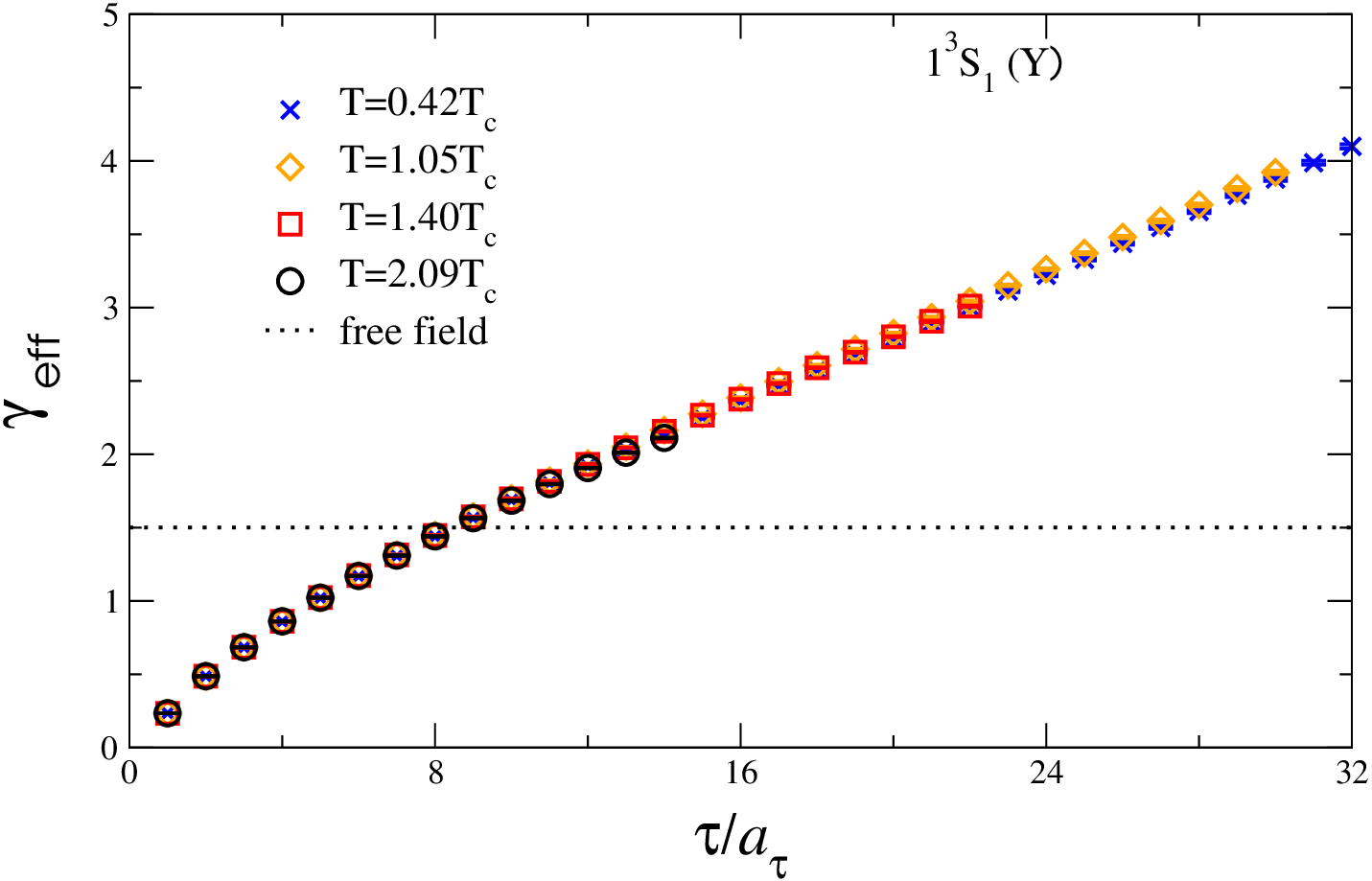}}
{\includegraphics[width=6.5cm]{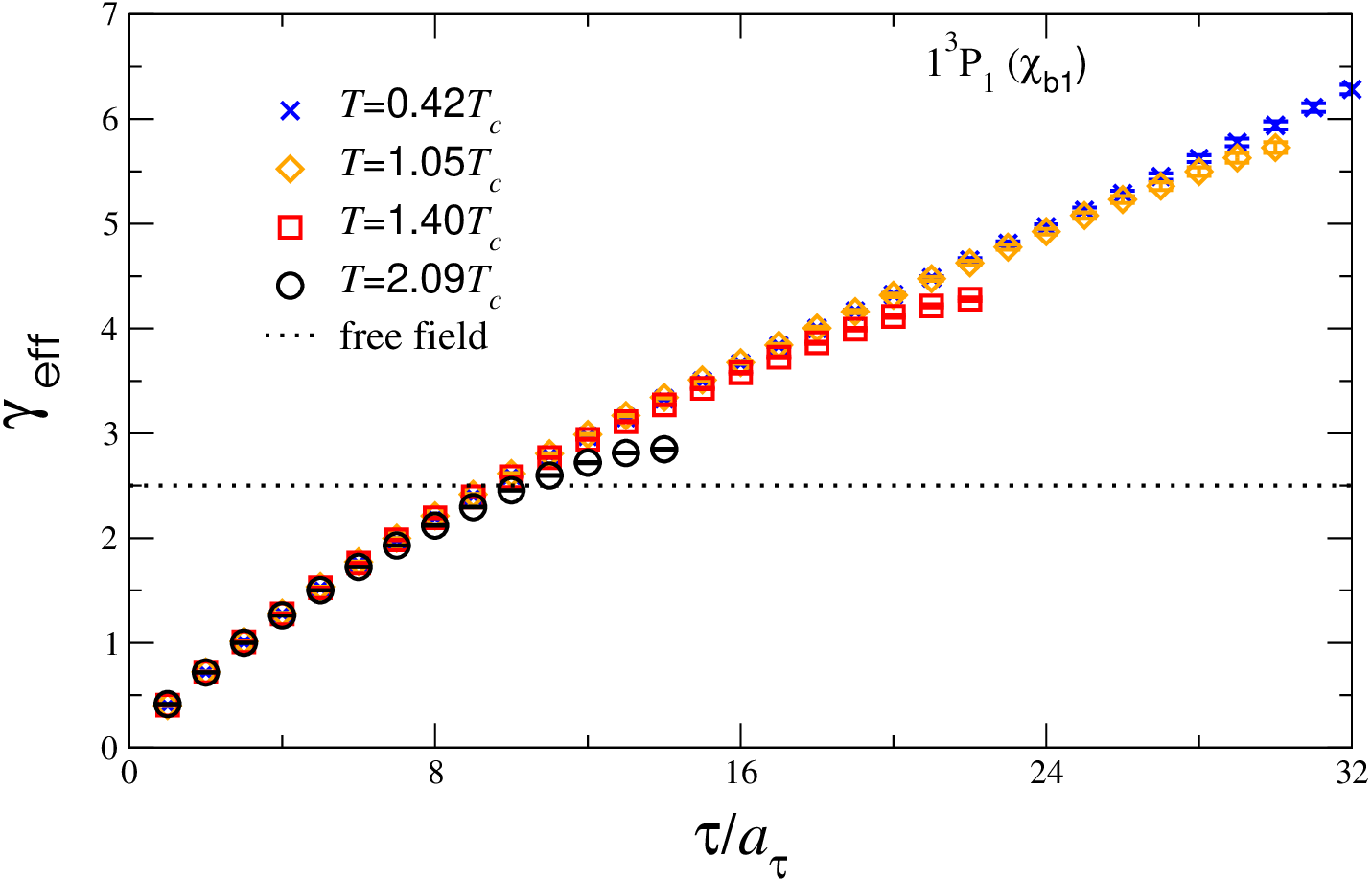}}
 \caption{
 Effective exponents $\gamma_{\rm eff}(\tau)$ for the $\Upsilon$ (left)
and $\chi_{b1}$ (right), as a function of Euclidean time for various
temperatures. The dotted line indicates the non-interacting 
result in the continuum (Ref.\ \cite{Aarts:2010ek}).} 
  
 \label{fig:power}
 \end{figure}

\section{ $\Upsilon$ and $\chi_b$  correlators in the plasma }

We discuss here the results one can obtain from the analysis of correlators
in Euclidean time, focusing on the $\Upsilon$ and $\chi_b$ 
states \cite{Aarts:2010ek}.

In Fig.\ 1, standard effective masses, defined by 
$
 m_{\rm eff}(\tau) = -\log[G(\tau)/G(\tau-a_\tau)],
$ are shown for both the $\Upsilon$ and the $\chi_{b1}$ propagators at 
various temperatures. Single exponential decay should yield a 
$\tau$-independent plateau. In both cases we find that at the lowest 
temperature, $T=0.42T_c$, exponential behaviour is visible provided one 
goes to late euclidean times. At higher temperature 
we can see  first indications that the $\Upsilon$ is not sensitive 
to the quark-gluon plasma up to $T\simeq 2T_c$, while the $\chi_b$ may 
melt at much lower temperatures, as exponential behaviour is no longer 
visible. 

To visualize the approach to quasi-free behaviour 
we construct effective power plots \cite{Aarts:2010ek}, using the definition 
$ \gamma_{\rm eff}(\tau) = -\tau\frac{G'(\tau)}{G(\tau)}
 = -\tau\frac{G(\tau+a_\tau) - G(\tau -a_\tau)}{2a_\tau G(\tau)},
$
 where the prime denotes the (discretized) derivative.  For a power decay, 
$G(\tau)\sim \tau^{-\gamma}$, this yields a constant result, $\gamma_{\rm 
eff}(\tau)=\gamma$.  The results are shown in Fig.\ 2. 
We confirm again that the $\Upsilon$ displays a very mild temperature 
dependence, while for the $\chi_{b1}$ the effective power tends
to flatten 
out. As discussed, in the continuum limit and in the free case one
should observe a flat behaviour. The observed flattening might thus
suggest an approach to a free field behaviour, however several caveats
apply. 
Also shown are the effective exponents in the continuum non-interacting limit. 
In the case of the $\chi_{b1}$, we observe that the 
effective exponent tends towards the non-interacting result at the highest 
temperature we consider.

To extend and validate these results, we proceed to the calculation
of the spectral functions. 

\section{Spectral functions}

\begin{figure}[!t]
\begin{center}
  \includegraphics*[width=12cm]{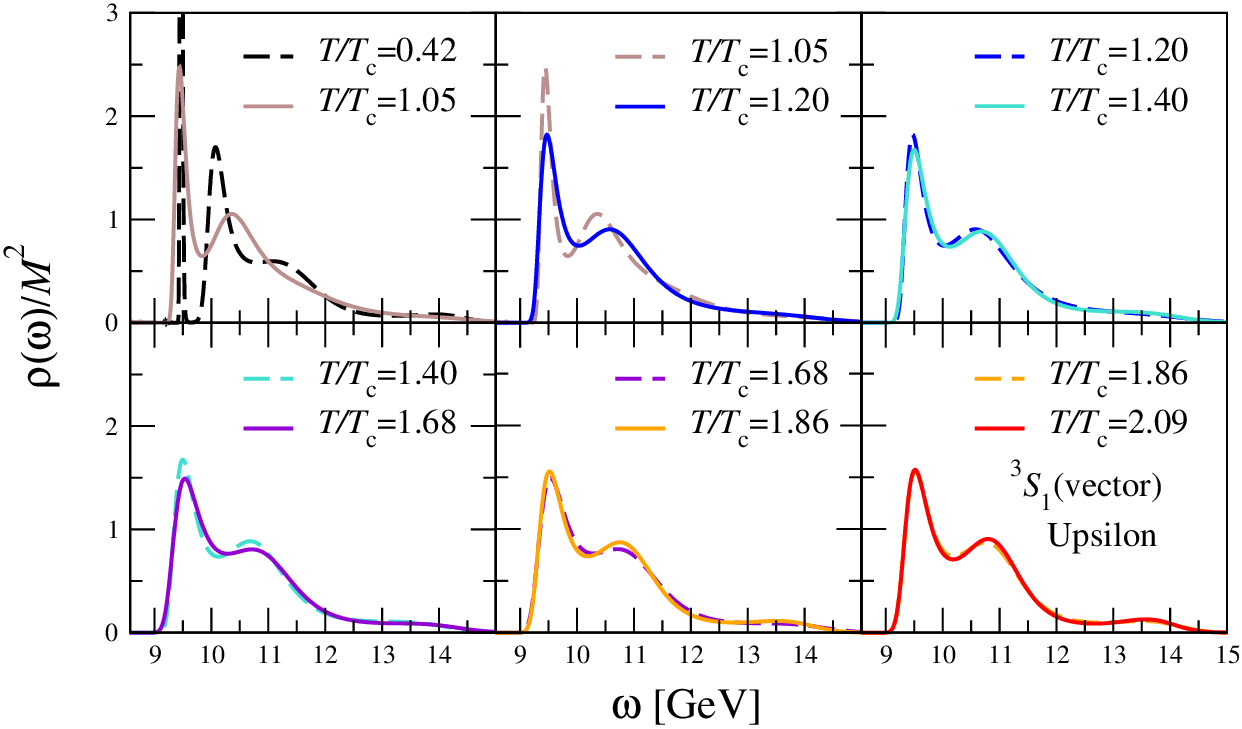} 
  \includegraphics*[width=12cm]{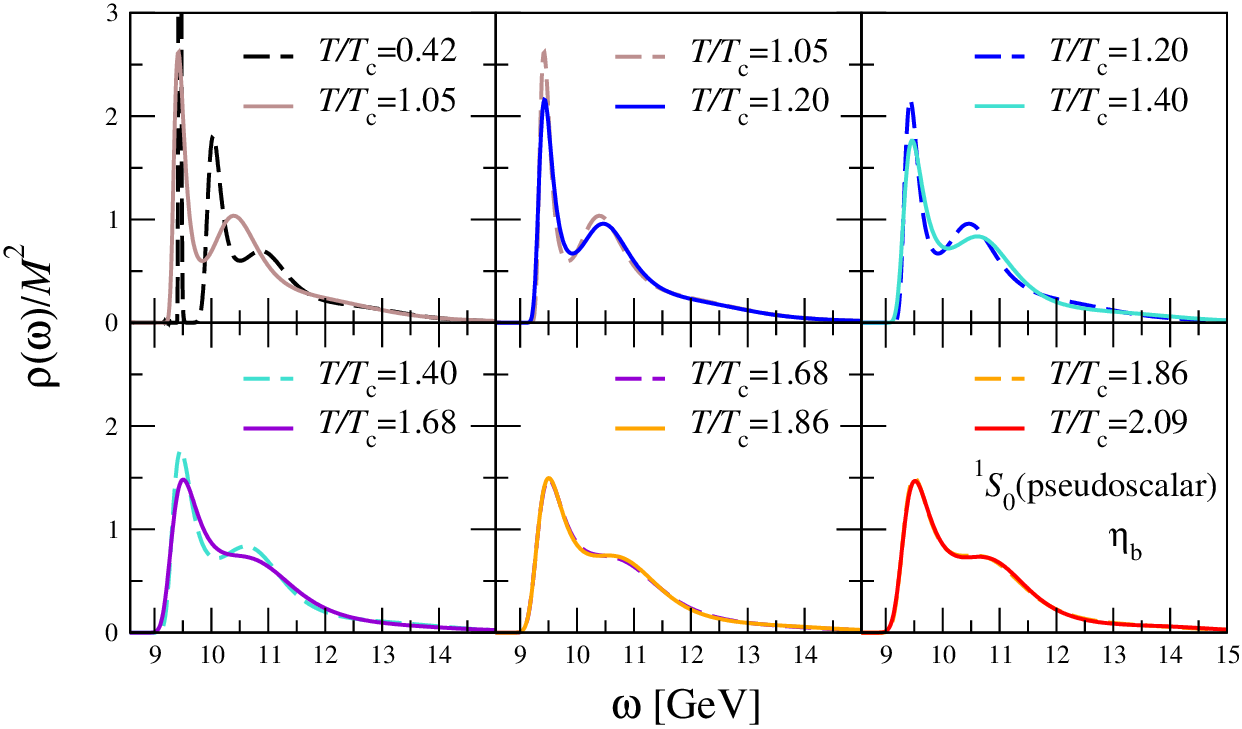} 
 \end{center}
 \caption{ Spectral functions $\rho(\omega)$, normalised with the
heavy quark mass, in the vector ($\Upsilon$) channel 
 (upper panel) 
and in the pseudoscalar ($\eta_b$) channel (lower panel)
for all temperature available. The subpanels are ordered from cold (top left) to
hot (bottom right). Every subpanel contains two adjacent temperatures to
facilitate the comparison (Ref.\ \cite{Aarts:2011sm}).  }
\label{fig:rho_all_upsilon}
\end{figure}

 Our main results for the spectral functions \cite{Aarts:2011sm}
can be seen in Fig.\ \ref{fig:rho_all_upsilon}, which shows that as the temperature is increased
the ground state peaks of the $\Upsilon$ and $\eta_b$ remain visible.
One caveat applies also here: the apparent width at
zero temperature is most likely due to a lattice/MEM artifact, and this
calls for an analysis of the discretization effects on the spectral function,
which is one of our ongoing projects. 
The peaks associated to  the  excited
states become suppressed at higher temperature and are no longer
discernible at $T/T_c\sim 1.68$.  The temperature dependence of the
position and width of the ground state peaks is compared to analytical predictions
obtained within the EFT formalism \cite{Brambilla:2010vq}.  
We note that the survival of the $\Upsilon(1S)$
state and suppression of excited states is consistent with the recent
experimental results \cite{:2012fr}.

\begin{figure}[!t]
\begin{center}
  \includegraphics*[height=4cm]{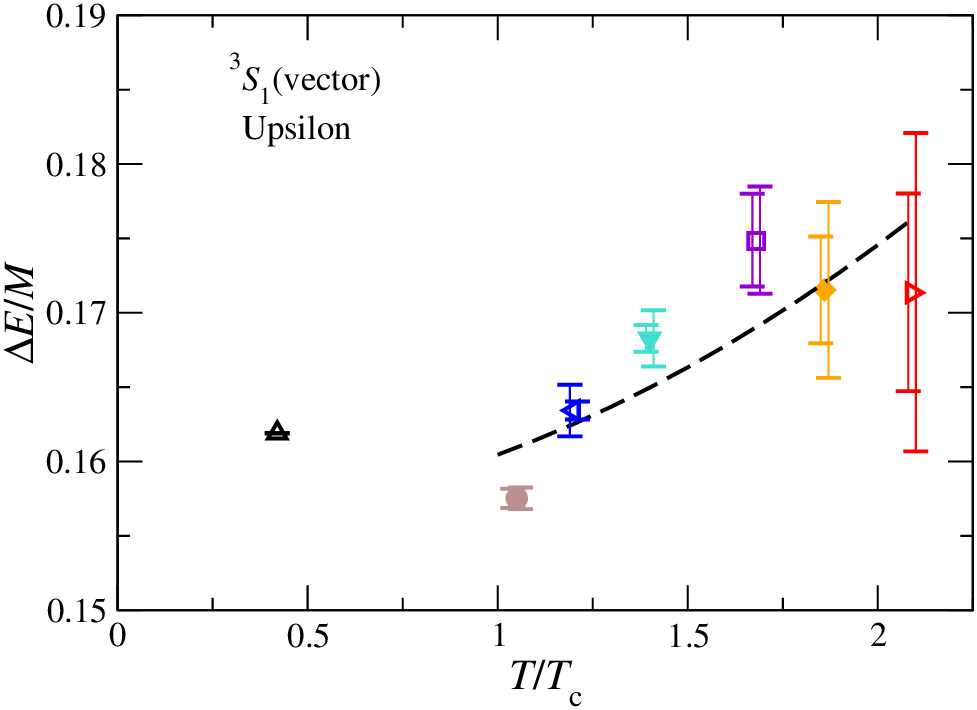}
  \includegraphics*[height=4cm]{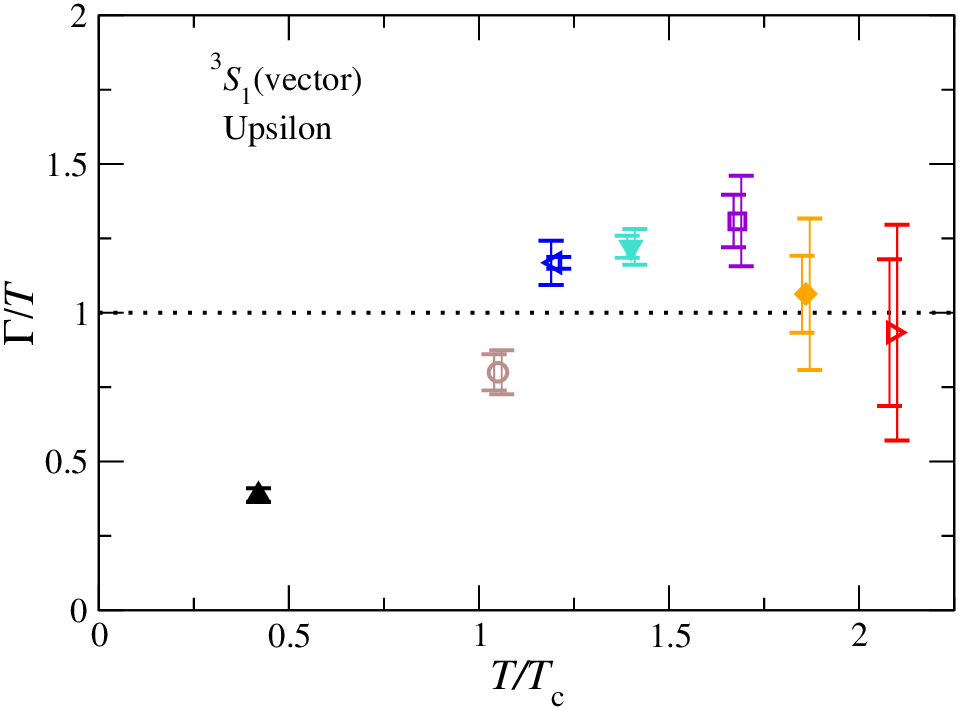}
 \end{center} 
 \caption{
Position of the ground state peak $\Delta E$, normalised by the
heavy quark mass (left), 
and the upper limit on the width of the ground state peak, normalised
by the temperature (right), 
as a function of $T/T_c$ in the vector ($\Upsilon$) channel.
Similar results have been obtained  for the
pseudoscalar ($\eta_b$) channel  (Ref. \cite{Aarts:2011sm}).
}
 \label{fig:mass}
\end{figure}

From these spectral functions we can determine the mass (from the peak
position) and (an upper bound on) the width of the ground state at each temperature.
In Fig.\ \ref{fig:mass} we show the temperature dependence of the mass
shift $\Delta E$, normalised by the heavy quark mass and the
temperature dependence of the width, normalised by the temperature.

We now  contrast our results with analytic
predictions derived assuming a weakly coupled plasma. 
According to Ref.\ \cite{Brambilla:2010vq}, the thermal contribution to the width is given, at leading order in the weak coupling and large mass expansion, by
\be
\frac{\Gamma}{T} =  \frac{1156}{81}\as^3 \simeq 14.27\as^3,
  \ee
  i.e., the width increases linearly with the temperature.
 If we take as an estimate from our results that  $\Gamma/T\sim 1$, we find that
 this corresponds to $\as\sim 0.4$, which is a reasonable result.
(It would be of interest to compute $\as$ directly on our configurations.)
In the same spirit  the thermal mass shift is given in Ref.~\cite{Brambilla:2010vq} by
\begin{equation} 
 \delta E_{\rm thermal} =  \frac{17\pi}{9}\as\frac{T^2}{M} \simeq
 5.93\as\frac{T^2}{M}. \label{eq:deltaE-pert}
\end{equation}
In these simulations we have $T_c\sim 220$ MeV, $M\sim 5$ GeV.  Taking
these values together with $\as\sim 0.4$ as determined
above, Eq.\ \eqref{eq:deltaE-pert} becomes
\be
 \frac{\delta E_{\rm thermal}}{M} =5.93\as \left(\frac{T_c}{M}\right)^2\left(\frac{T}{T_c}\right)^2 \sim 0.0046\left(\frac{T}{T_c}\right)^2.
 \ee
In order to contrast our results with this analytical prediction, we have compared the temperature dependence of the peak positions to the simple expression
\begin{equation}
 \frac{\Delta E}{M} =c+ 0.0046\left(\frac{T}{T_c}\right)^2, 
\end{equation}
where $c$ is a free parameter. This is shown by the dashed line in
Fig.\ \ref{fig:mass} (left panel). The numerical results and the
analytic ones are not inconsistent, within the large errors, which 
encourages us to pursue further these studies.

\section{Momentum Dependence}

We extend now our study of bottomonium spectral functions in the
quark-gluon plasma to nonzero momentum \cite{Aarts:2012ka}. 
Note that  effective theories analysis predict
significant momentum effects at large momenta, and that 
current CMS results have been obtained at
large momenta. So there is both phenomenological and experimental
motivation for these studies.

 The momenta and velocities that are accessible on the lattice are
 constrained by the discretization and the spatial lattice
 spacing. The lattice dispersion relation  reads
   \be
 a_s^2\pv^2 = 4\sum_{i=1}^3 \sin^2\frac{p_i}{2}, \quad\quad
   p_i = \frac{2\pi n_i}{N_s}, \quad\quad
    -\frac{N_s}{2} < n_i \leq \frac{N_s}{2}.
 \ee
   To avoid lattice artifacts, only momenta with $n_i<N_s/4$ are used: we consider the combinations (and permutations thereof) given in table \ref{tab:mom}.
   The largest momentum, using $\nv = (2,2,0)$, is $|\pv| \simeq 1.73$ GeV, corresponding to $v = |\pv|/M_S \simeq 0.2$. Therefore, the range of velocities we consider is non-relativistic.

\begin{table}[t]
\begin{center}
\begin{tabular}{| l | ccccccc | }
\hline
$\nv$ 			& (1,0,0)	& (1,1,0)   & (1,1,1)  &  (2,0,0)	& (2,1,0)	& (2,1,1)	& (2,2,0) 	\\
$|\pv|$ (GeV)		& 0.634    & 0.900 	& 1.10 	& 1.23	& 1.38 	& 1.52 	& 1.73	\\
$v$ [$\Upsilon(^3S_1)$]	& 0.0670 	& 0.0951	& 0.116	& 0.130	& 0.146	& 0.161 	& 0.183 	\\
$v$ [$\eta_b(^1S_0)$]	& 0.0672 	& 0.0954 	& 0.117 	& 0.130 	& 0.146 	& 0.161 	& 0.183	\\
\hline
\end{tabular}
\vspace*{0.2cm}
 \caption{Nonzero momenta used in this study. Also indicated are the
   corresponding velocities $v=|\pv|/M_S$ of the ground states in the
   vector ($\Upsilon$) and pseudoscalar ($\eta_b$) channels, using the
   ground state masses determined previously  \cite{Aarts:2010ek},  $M_{\Upsilon}=9.460$ GeV and  $M_{\eta_b}=9.438$ GeV.
 }  
\label{tab:mom}
\end{center}
\end{table}

\begin{figure}[h]
\begin{center}
\epsfig{figure=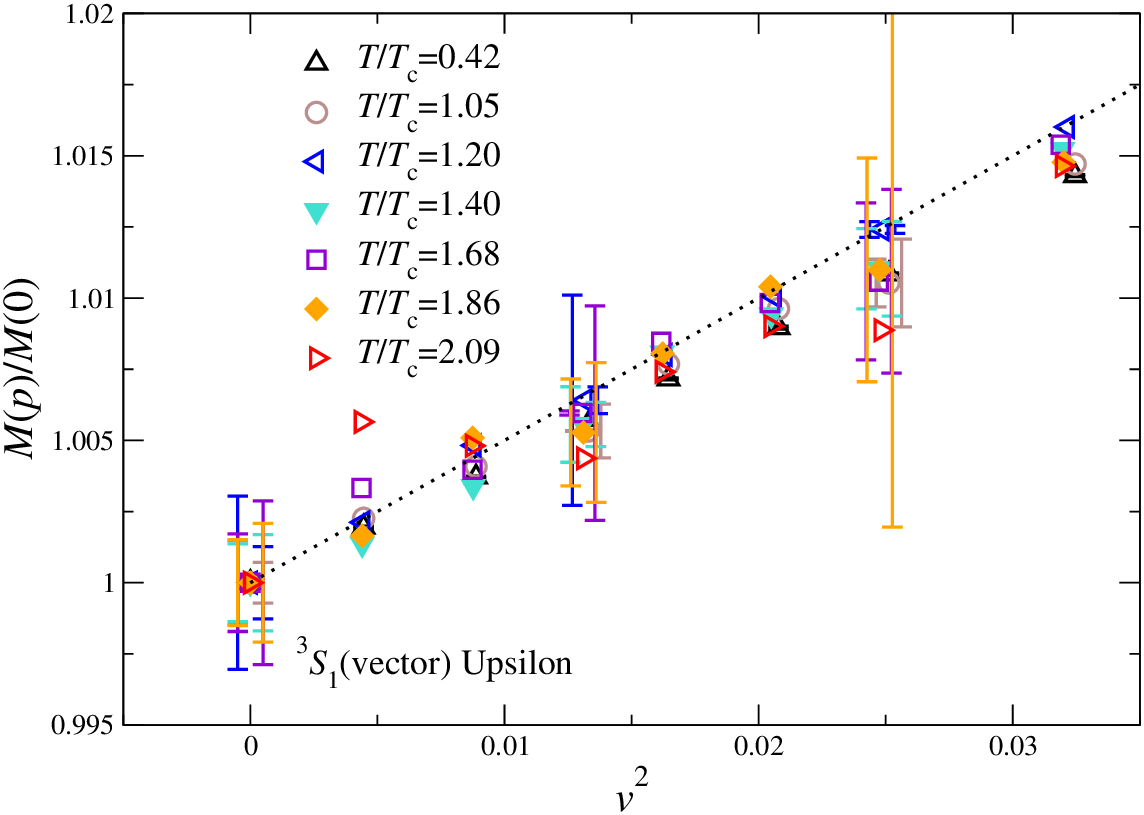,width=0.4\textwidth}
\epsfig{figure=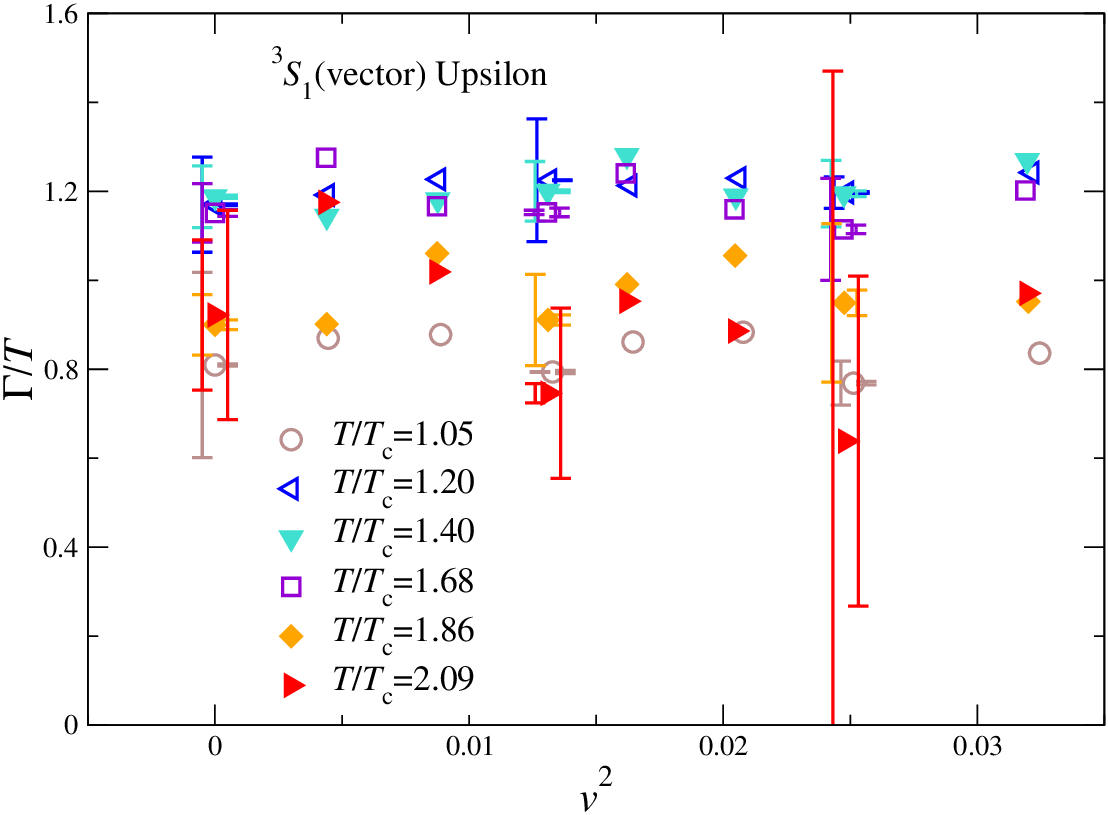,width=0.4\textwidth}

\end{center}
 \caption{
Position of the ground state peak $M(\pv)/M(0)$ (left)
and  the upper limit on the width of the ground state peak, normalized
by the temperature, 
$\Gamma/T$ (right), as a function of the velocity squared ($v^2$) in the vector ($\Upsilon$) channel %(above). 
Analogous results have been obtained for the 
 the pseudoscalar ($\eta_b$) channel. 
The dotted line in the left figure represents 
$M(\pv)/M(0)=1+\half v^2$ (Ref.\ \cite{Aarts:2012ka}).  }
 \label{fig:mass-width}
\end{figure}

As in the zero momentum case, we extract masses and widths from our
correlators using MEM, and we display them in Fig. 5.   
We observe  that the peak position increases linearly with $v^2$, as expected. Assuming the lowest-order, non-relativistic expression $M(\pv)=M(0) + \pv^2/2M(0)$, one finds
\be
\frac{M(\pv)}{M(0)} = 1+ \frac{\pv^2}{2M^2(0)} = 1+\half v^2,
\ee
which is indicated with the dotted lines in the left figures.

The dependence on the velocity  can be compared with EFT predictions. In Ref.\ \cite{Escobedo:2011ie} a study of the velocity dependence was carried out  in the context of QED, working in the rest frame of the bound state (i.e.\ the heat bath is moving). In order to compare with our setup, we consider the case in which the temperature is low enough for bound states to be present and that the velocities are non-relativistic. In that case, one finds \cite{Escobedo:2011ie}, in the rest frame of the bound state and at leading order in the EFT expansion, 
\be
\frac{\Gamma_v}{\Gamma_0}  = \frac{\sqrt{1-v^2}}{2v}\log\left(\frac{1+v}{1-v}\right),
\ee
where $\Gamma_0$ is the width at  rest.
Interpreting the width as an inverse lifetime, one can express this
result  in the rest frame of the heat bath by dividing with the Lorentz factor $\gamma = 1/\sqrt{1-v^2}$. An expansion for non-relativistic velocities then yields
\be
\frac{\Gamma_v}{\Gamma_0} =   1 -\frac{2v^2}{3} +  {\cal O}\left(v^4\right).
\ee
If we apply this result  to our study of bottomonium,  we find that 
the effect of the nonzero velocity shows up as a correction at the percent level (recall that $v^2\lesssim 0.04$), which is beyond our level of precision but consistent with the observed $v$ independence within errors.
Similarly, additional thermal effects in the dispersion relation are currently beyond our level of precision.
In summary, the observation in our low-momentum range
are consistent with Ref.\ \cite{Escobedo:2011ie}, and in 
order to observe the predicted  
non-trivial momentum dependence we need to explore  larger momenta.

\section{Summary }
\label{sec:sum}

We have presented our results on bottomonium in the quark-gluon plasma,
for temperatures up to  $2.1T_c$, at the threshold of the region
 currently  explored by LHC heavy-ions experiments. 
Our analysis uses  full relativistic
gauge field configurations, and a non-relativistic approach for the
bottom quarks. Our goal is to reach a quantitative, systematic control
on the results, which is called for by the new level of experimental
accuracy. Related to this, we aim at understanding in detail the range
of validity of the available effective models. This is a task of foremost importance,
since an accurate analytic modeling is needed as an input to real time
evolution. 

Our results so far were based on a set of gauge field configurations 
obtained with two flavors of dynamical quarks.
We have used anisotropic lattices in order to maximize
the number of points in Euclidean time --- a much needed feature for a successful
MEM analysis. In this way even on our hottest lattices ($T=2.1T_c$)
we have been able to analyze correlators and proceed to the calculation
of the spectral functions. 

On these lattice ensembles 
we have observed that the temperature affects only very mildly the 
$\Upsilon$ and $\eta_b$ correlators, suggesting the survival of the fundamental
states in these channels 
up to $T \simeq 2 T_c$. A subsequent calculation of the spectral 
functions carried out with a MEM analysis has confirmed
the survival of the fundamental states and  
revealed the melting of the excited states in the same channels 
above $1.4 T_c$. The 
correlators for the $\chi_b$ and $h_b$  indicate a quick melting of these 
states above $T_c$.

On the methodological side, 
we have argued that the peculiar nature of non-relativistic 
dynamics greatly simplifies the MEM analysis and enhances its reliability.
We  have invested a considerable effort in monitoring the systematic
effects associated with the MEM analysis, which is described in our papers,
and we feel that we have them well under control. The special form
of the NRQCD spectral relations shows that the NRQCD spectral function is
an inverse Laplace transform of the correlator. In a nutshell, this 
makes the MEM analysis easier than in the full relativistic case, and perhaps
opens the way to a direct, more simple evaluation of the spectral functions
themselves.

Our activities in the near future are evolving along different lines.
Firstly,  new results which will be available in the near future
include a MEM analysis
for the $\chi_b$ and $h_b$ channels.

Secondly, we wish to approach more closely the continuum limit, with 
a physical matter content. To this end, simulations with a reduced
lattice spacing, and the inclusion of the strange quarks, are
already in progress. On the NRQCD side, on the new $N_f = 2+1$ lattices 
the quark mass is tuned precisely using the kinetic mass
and the lattice NRQCD dispersion relation. Moreover 
we are considering  using the   1-loop determinations of the 
$\sigma\cdot B$ term in the 
NRQCD coefficients   which has recently produced an
impressive improvement in the determination of the (zero temperature)
hyperfine splitting \cite{Hammant:2011bt} . 
All this will enhance the control over various lattice 
systematic errors. In a longer term perspective, relativistic beauty
might also become feasible on these finer lattices, mirroring the
approach that is already being followed for charmonium
\cite{Aarts:2007pk,charm-new}.  This will provide a 
definitive check
of the NRQCD approach. 

Next, as for the momentum dependence, we have mentioned that 
both experimental and phenomenological activities call
for calculations at larger momenta: a sizable dependence is
expected at large momenta and large momenta are those 
studied by CMS.  Large momentum studies  will be possible
on our finer lattices and this is being planned. 

As a final remark, leading to further developments, 
an important aspect of our study is the comparison with
effective theories. Within our limited precision, and limited explored
range, the comparison has been satisfactory. This motivates us
to enhance this analysis  in the future. To carry out this
program, besides the improvement already described, we will have
to explore a wider set of parameters.  
One economic way of doing this is to vary the heavy quark mass, studying
artificial mesons:  this  has been very 
helpful at zero temperature in cross checking lattice results
and effective model predictions --- for instance
the results for the light meson spectrum are being compared with
chiral perturbation theory and the heavy meson spectrum with
heavy quark effective theory. In the same spirit  
we would like to know  the dependence of the bottomonium spectrum
on the heavy quark mass, and temperature, and compare it  with
various effective models in their respective ranges of applicability.
Remember in fact that different effective models hold true
in correspondence with different relative values of the relevant
scales such as
$M, M\as, T, M\as^2, m_D$, so upon varying mass
and temperatures one should observe a crossover between different behaviours,
thus building confidence in using the corresponding models. 
This program has started and its first, preliminary
results have been reported at  Lattice  2012\cite{Kim:2012by}.

\ack
We wish
to thank the Organisers for their suggestion that MPL  replace the intended 
invited speaker --- S.M. Ryan ---  who had to cancel her participation, 
so that our work could still be presented at the meeting.

%%%%%%    
\end{document}